\documentclass[reprint, amsmath,amssymb, aps, prd]{revtex4-2}

\usepackage{graphicx}
\usepackage{dcolumn}
\usepackage{bm}
\usepackage{upgreek}
\usepackage{hyperref}

\usepackage{pennames}

\usepackage{xstring}
\makeatletter
\AtBeginDocument{%
  \def\doi#1{\href{https://doi.org/#1}{doi:\StrSubstitute{#1}{_}{\_}}}}
\makeatother

\begin{document}


\title{A rotation-equivariant graph neural network for learning hadronic SMEFT effects}

\thanks{Prepared for submission to PRD.}%

\author{Suman Chatterjee}
\email{suman.chatterjee@oeaw.ac.at}
\affiliation{Institute for High Energy Physics, Austrian Academy of Sciences, Nikolsdorfergasse 18, A-1050 Vienna, Austria}

\author{Sergio S\'anchez Cruz}
\email{sergio.sanchez.cruz@cern.ch}
\altaffiliation[Now at the ]{European Organization for Nuclear Research (CERN)}
\affiliation{Department of Physics, University of Z\"urich, Winterthurerstrasse 190, 8057, Z\"urich, Switzerland}

\author{Robert Sch\"ofbeck}
\email{robert.schoefbeck@oeaw.ac.at}
\affiliation{Institute for High Energy Physics, Austrian Academy of Sciences, Nikolsdorfergasse 18, A-1050 Vienna, Austria}

\author{Dennis Schwarz}
\email{dennis.schwarz@oeaw.ac.at}

\affiliation{Institute for High Energy Physics, Austrian Academy of Sciences, Nikolsdorfergasse 18, A-1050 Vienna, Austria}

\date{\today}

\begin{abstract}
We introduce a graph neural network architecture designed to extract novel phenomena in the Standard Model Effective Field Theory (SMEFT) context from LHC collision data. The proposed infrared- and collinear-safe architecture is sensitive to the angular orientation of radiation patterns in jets from hadronic decays of highly energetic massive particles. Equivariance with respect to rotations around the jet axis allows for extracting the information on the angular orientation decoupled from the jet substructure. We demonstrate the robustness of the approach and its potential for future probes of the SMEFT at the LHC through toy studies and with realistic event simulations of the WZ process in the semileptonic decay channel.
\end{abstract}

\maketitle


\section{Introduction}\label{sec:intro}

The Large Hadron Collider (LHC) is a veritable gold mine of data, among whose most complex signatures are collimated sprays of particles~(jets) originating from the hadronic decays of boosted massive particles. This richness presents a challenge and a fertile ground for exploring fundamental physics in hadronic final states.

In recent years, machine learning algorithms have been increasingly adapted to more complex data representations, with the ensuing rise of the input feature dimension tamed by imposed symmetries of the underlying physical problem~(see Refs.\cite{Carleo:2019ptp,Erdmann:2021jbm,Tanaka2021} for an overview). In particular, message-passing graph neural networks (gNNs)~\cite{Kipf:2016gmz,DBLP:journals/corr/GilmerSRVD17,Wang:2018nkf} learn from relationships among particles that are interpreted as point clouds with manifest permutation invariance~\cite{Erdmann:2018shi, Qu:2019gqs,Onyisi:2022hdh,DeZoort:2023vrm}.
Such developments have markedly improved the tagging performance for the various objects that are reconstructed as highly energetic jets at collider experiments~\cite{Kasieczka:2019dbj}. 
Incorporating invariance or equivariance for symmetry groups specific to certain tagging challenges refines the algorithms' inductive bias and is an active area of research~\cite{Satorras2021EnEG,Villar:2021wnx,Dolan:2020qkr,Favoni:2020reg,Bulusu:2021njs,Gong:2022lye,Bogatskiy:2022hub,Favoni:2022mcg,Bogatskiy:2022czk,Buhmann:2023pmh,Murnane:2023kfm,Bogatskiy:2023nnw}. 
In particular, infrared and collinear (IRC) safety was integrated into gNNs by means of energy-weighted message-passing~\cite{Konar:2021zdg}, generalizing Energy Flow Networks~\cite{Komiske:2018cqr},  and providing robustness to uncertainties in the modeling of the splitting or merging of soft and collinear particles.
A rich technological toolkit thus extends the experiments' grasp on lower-level data representations that are particularly relevant for the hadronic final states considered in this work.

In the theoretical domain, a parallel interest in more low-level representations of hypothetical new phenomena is fueled by the absence of any compelling signal of new resonant phenomena. The Standard Model Effective Field Theory (SMEFT)~\cite{Buchmuller:1985jz,Giudice:2007fh,Grzadkowski:2010es} has emerged as the preferred language for describing hypothetical phenomena below an assumed energy scale, that separates the energy scale of the LHC from resonant phenomena at much higher scales.
The SMEFT is a powerful framework extending the standard model~(SM) Lagrangian by field monomials whose prefactors, the Wilson coefficients, are the parameters of interest (POIs) in experimental measurements. These give the experimentalist a low-level representation of a wide range of hypothetical high-scale models.

The SMEFT organizing principle is the operators' mass dimension, starting at six for the new physics scenarios relevant at the LHC~\cite{Grzadkowski:2010es}. Because the lowest-order amplitude modifications are linear in the Wilson coefficients, the cross-section deviations are accurately described by quadratic polynomials within the SMEFT's range of validity. This simple analytic structure, in turn, is exploited in various simulation-based inference techniques~\cite{Cranmer:2015bka,Brehmer:2018kdj,Brehmer:2018eca,Brehmer:2018hga,Brehmer:2019xox,Brehmer:2019gmn,Butter:2021rvz,Chatterjee:2022oco,GomezAmbrosio:2022mpm,Chen:2020mev,Chen:2023ind}.

The linear term in the polynomial describes the SM--SMEFT interference.  It is the only contribution not subject to further contributions from SMEFT operators with a higher mass dimension, and, therefore, the unambiguous harbinger of \mbox{dimension-6} SMEFT effects. In a wide range of final states involving decays of massive vector bosons, however, a naive experimental analysis may unintentionally remove the sensitivity to the linear contribution. In these cases, a dedicated angular analysis can sometimes ``resurrect'' the interference terms~\cite{Panico:2017frx,Banerjee:2019twi,Banerjee:2020vtm,BuarqueFranzosi:2021wrv,Degrande:2021zpv}. In such cases, the orientation of the decay planes of the \PW~or \PZ~boson provides crucial sensitivity because it can resolve the amplitudes' helicity configuration which is altered in the SMEFT~\cite{Panico:2017frx}. Analyses of leptonic final states of diboson processes~(VV with V=\PW, \PZ, \PG, or H) already profit from the boost in experimental sensitivity~\cite {CMS:2021icx,CMS:2021cxr,ATL-PHYS-PUB-2021-022} and served as a motivating use-case for the development of machine-learned optimal observables~\cite{Chen:2020mev,Chatterjee:2022oco,GomezAmbrosio:2022mpm,Chen:2023ind}.
\begin{figure}
\includegraphics{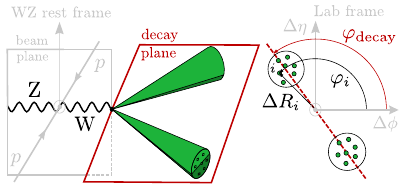}
\caption{\label{fig:decay} Sketch of the  $\Pp\Pp\rightarrow\PW\PZ$ process, the decay plane, the decay plane angle $\varphi_{\textrm{decay}}$ of the $\PW$~boson decay. The $\Delta R_i$ and $\varphi_i$ coordinates of a constituent $i$ from a highly energetic jet are defined with respect to the beam plane and the jet axis. }
\end{figure}

The main idea of this paper is to extend this simulation-based inference approach to hadronic final states and extract SMEFT sensitivity via a gNN that is equivariant with respect to azimuthal rotations around the axis of a highly energetic jet, originating from the hadronic decay of a boosted massive particle. The variable-length set of the jet's constituents is the input to the gNN. Its output is fed into dense layers that can also accept other features of the event.
The algorithm is tailored towards the linear SM--SMEFT interference and exploits the spatial angular orientation of the radiation patterns with respect to the event remainder. A prototypical situation is sketched in Fig.~\ref{fig:decay}.

The rest of the paper is organized as follows. In Sec.~\ref{sec:WZEFT} we describe the subject of our study, the semileptonic $\PW\PZ$ process, and the data sets' structure. In Sec.\ref{sec:algo}, we formulate the algorithm. We elucidate its main features in simple toy studies presented in Sec.~\ref{sec:toy}. The application to simulated $\PW\PZ$ events is provided in Sec.~\ref{sec:WZEFT-results}, and we give conclusions and an outlook in Sec.~\ref{sec:outlook}.



\section{The semileptonic \PW\PZ~final state}\label{sec:WZEFT}
Despite the relatively small number of expected events in the high-$p_{T}$ regime, we can nevertheless probe SMEFT operators as they induce energy-growing corrections to the SM amplitudes~\cite{Chen:2020mev,Chen:2023ind}.
Among the diboson final states providing SMEFT sensitivity to LHC data analyses~\cite{CMS:2021icx,CMS:2021cxr,ATL-PHYS-PUB-2021-022}, we chose the comparably simple case of $\Pp\Pp\rightarrow \PW\PZ$ as a motivating example.
We consider the semileptonic decay channel, where
$\PZ\rightarrow \bar\ell\ell$ and $\PW\rightarrow \bar \Pq \Pq$, and restrict to the highly energetic phase space $p_{\textrm{T}}(\PW)>300$~GeV, where the \PW~boson is reconstructed as a  jet by the anti-$k_{\textrm{T}}$~\cite{Cacciari:2008gp} algorithm in the \texttt{FASTJET} implementation~\cite{Cacciari:2011ma} with a radius parameter of $R=0.8$~(AK8) of approximately massless jet-constituents. 
We focus on the interference contribution to the differential cross-section from the operators
\begin{align}
\mathcal{O}_{\PW}&=\varepsilon^{ijk}W_\mu^{i\nu}W_\nu^{j\rho}W_\rho^{i\mu}\quad\textrm{and}\nonumber\\
\mathcal{O}_{\widetilde \PW}&=\varepsilon^{ijk}\widetilde W_\mu^{i\nu} W_\nu^{j\rho}W_\rho^{i\mu}
\end{align}
whose Wilson coefficients we denote by $C_{\PW}$ and $C_{\widetilde\PW}$. 
The operators $\mathcal{O}_{\PW}$ and $\mathcal{O}_{\widetilde \PW}$ induce CP-even and CP-odd modifications of the gauge boson self-interactions, while another operator, $\mathcal{O}_{\PH\Pq}^{(3)}=(H^\dagger i\overset\leftrightarrow{D}{}^i_\mu H)\left(\bar q_L\sigma^i\gamma^\mu q_L\right)$, modifies the light quark--gauge boson vector-like interaction.
At the linear level, these operators exhibit energy growth with the center-of-mass energy of the diboson system, denoted by $s$. 
 The operator $\mathcal{O}_{\PH\Pq}^{(3)}$ dominantly contributes to the helicity configuration where both bosons are longitudinally polarized and where SM contribution is constant in $s$ at high energy. Therefore, the SM--SMEFT interference term introduces energy growth but does not modify the distributions of the azimuthal decay-plane angles. Therefore, we do not consider it further.

 The operators $\mathcal{O}_{\PW}$ and $\mathcal{O}_{\widetilde \PW}$, in contrast, contribute energy-growing amplitudes for transverse gauge boson polarisation with the same helicities which are small in the SM. Consequently, the interference contribution vanishes in the tail if it is not resurrected by a dedicated angular analysis~\cite{Panico:2017frx,Chen:2020mev,Chen:2023ind}. In the leptonic decay channels, the azimuthal orientation of the decay plane $\varphi_{\textrm{decay}}$, as shown in Fig.~\ref{fig:decay}, drives the sensitivity. A dedicated multivariate analysis of the fully leptonic decay mode exploits this fact by extracting SMEFT sensitivity from global event kinematics, i.e., per-event features that are reconstructed from lepton kinematics and in particular angular observables in the diboson rest frame~\cite{Chen:2020mev,Chen:2023ind}. 

In our case, we attempt to similarly access this sensitivity by extracting it with a graph neural network. Because of the special role of the azimuthal decay plane angle, we construct it equivariantly with respect to rotations of the boosted jet's constituents around the jet axis. The relevant symmetry group is SO$(2)$, irrespective of whether we choose our reference frame as the lab frame or, as done in Fig.~\ref{fig:decay}, as the rest frame of the $\PW\PZ$ system.  Our events are instances in a data set 
\begin{equation}
\mathcal{D}=\big\{\boldsymbol{x}_{\textrm{global},j},\{\boldsymbol{x}_{\textrm{p}}\}_{i=1}^{N_p(j)}\big\}{}_{j=1}^{N_{\textrm{events}}}\label{eq:dataset}
\end{equation}
where $\boldsymbol{x}_{\textrm{global}}$ denotes optional global event features, pertaining to the kinematics of the $\PW\PZ$~candidate event, the $\PZ$ boson, the transverse missing energy, etc. In each event, there is also a list of $N_{p}$ approximately massless constituent particles of an AK8~jet~\cite{Cacciari:2008gp}. These constituents could, for example, be provided by a particle-flow algorithm~\cite{CMS:2017yfk}. Each particle's feature vector $\boldsymbol{x}_{\textrm{p}}$ contains the four-momentum $p^{\mu}_i$ and, in principle, a number of features which could, e.g., represent the quality of the particle tracks association with the primary collision vertex, its probability of originating from a pileup vertex~\cite{Bertolini:2014bba}, or the charge of the particle. We emphasize that the data set in Eq.~\ref{eq:dataset} is hierarchical in the sense that the reference frame for the $p^\mu_i$ depends on the event kinematic that also enters via $\boldsymbol{x}_{\textrm{global}}$. There are two natural choices for these reference frames. In the lab-frame, the constituents' four-momentum is given by the transverse momentum $p_{\textrm{T},i}$ of the particle and the azimuthal angular difference to the jet axis $\Delta \phi_i$, as well as the difference in pseudo-rapidity $\Delta\eta_i$. Alternatively, we can obtain a polar and an azimuthal angle after a Lorentz boost into the diboson rest frame. In this case, we can align the z-axis of the spherical coordinate system with the jet axis and measure the azimuthal angle, denoted by $\varphi_i$, with respect to the beam plane as depicted in Fig.~\ref{fig:decay}. 
In both cases, we denote Euclidean distances in the two-dimensional angular coordinates by $\Delta R$.
Either way, the data set encodes information on the per-event decay plane that is disguised in the radiation patterns mapped to the variable-length constituent vector. The algorithm described in the next section is designed to extract its SMEFT sensitivity.

\section{The algorithm}\label{sec:algo}
The constituents $\{\boldsymbol{x_p}\}_{i=1}^{N_{\textrm{p}}}$ of the highly energetic jet are individually reconstructed particles that can be viewed as a point cloud whose elements are feature vectors, composed of observables like four-momenta, charge, and other constituent properties.

General message-passing gNNs construct graphs on the point cloud and iteratively change the representations of the graphs' nodes by updating the feature vectors $\mathbf{h}_i$ by an aggregate of messages from the nodes in a neighborhood $\mathcal{N}(i)$ and,  possibly,  from the edges connecting the nodes~\cite{Wang:2018nkf,Kipf:2016gmz,DBLP:journals/corr/GilmerSRVD17}.  
In the most general case, a node update function $\psi^{(l)}$ and a message passing function $\phi^{(l)}$ are highly expressive learnable functions, typically implemented as multilayer perceptrons~(MLP). At each iteration $l+1$, these determine the feature vector of a node $i$ from the messages $^i\mathbf{m}_j^{(l)}=f_{\textrm{m}}^{(l)}(\mathbf{h}_i^{(l)}, \mathbf{h}_j^{(l)},\mathbf{e}_{ij})$ that is obtained from the node features $\mathbf{h}^{(l)}_{i}$, the neighbors' node features $\mathbf{h}^{(l)}_{j}$, and, possibly, from edge features $\mathbf{e}_{ij}^{(l)}$ via the general node-update formula
$\mathbf{h}_i^{(l+1)}=f_{\textrm{r}}^{(l)}\left(\mathbf{h}_i^{(l)},\Box_{j\in\mathcal{N}(i)}\,{}^i\mathbf{m}_j^{(l)}\right)$.  The aggregation function,  denoted by $\Box$,  is permutation invariant and accumulates the messages from a suitably defined neighbourhood $\mathcal{N}(i)$ of particle $i$.  After a number of $L$ iterations,  the nodes are read out by accumulating the features $\mathbf{h}^{(L)}$ with an aggregation function that typically spans all nodes in the graph. In the simplest setting, this output is fed into a final global read-out MLP whose parameters, together with the parameters of the message-passing and per-iteration read-out MLPs, $f_{\textrm{m}}^{(l)}$ and $f_{\textrm{r}}^{(l)}$, are adjusted to minimize a problem-specific loss function.

First,  we simplify the problem by demanding IRC safety via energy-weighted message passing, following Ref.~\cite{Konar:2021zdg}.  This requirement imposes a number of restrictions.  
The input features for iteration $l=0$ are only the four-vectors $p_i$ of the particles.  
The message passing function, specifically, is given by $^i\mathbf{m}_j^{(l)}=\omega_j^{(\mathcal{N}(i))}f^{(l)}_{\textrm{m}}(\hat p_i,\hat p_j)$ where $\hat p$ denotes the direction of the three-momentum of the particles.  The energy weighting is implemented via the relative hardness
\begin{align}
\omega^{\mathcal{N}}_j&=\frac{p_{\textrm{T},j}}{\sum_{k\in\mathcal{N}}p_{\textrm{T},k}}\label{Eq:energy-weighted-message}
\end{align}
where instead of the transverse particle momentum $p_{T}$,  we could alternatively use another measure of the hardness of the particle,  e.g., the energy or the three-momentum measured in a well-defined rest-frame.  Furthermore,  the definition of the neighborhood cannot depend on the occurrence of infrared or collinear splits of the particles,  excluding,  e.g., the otherwise common kNN algorithm.  
Instead,  a particle $j$ is defined to be in the neighborhood $\mathcal{N}(i)$ of a particle $i$ according to the Euclidean distance in the (pseudo-)rapidity--azimuth plane, \mbox{$\Delta R_{ij}=\sqrt{\Delta \phi(\hat p_i, \hat p_j)^2 + \Delta \eta(\hat p_i, \hat p_j)^2}\leq\Delta R$}. 
The threshold $\Delta R$ is a hyperparameter of the network.
Choosing a specific per-event reference frame for the $N_p$ massless constituents of a given W jet candidate, the gNN inputs for $l=0$ are represented by a feature list  $\{p_{T,i},\varphi_i,\Delta R_i\}_{i=1}^{N_{p}}$ as shown in Fig.~\ref{fig:decay}.

Our algorithm is also  SO$(2)$ equivariant.  In general,  a function $\gamma:X\rightarrow Y$ is equivariant with respect to group transformations $g$ of a group $G$,  if there are two  representations $T_g$ and $S_g$, acting on the spaces $X$ and $Y$, respectively, and the function satisfies $S_g(\gamma(x))=\gamma(T_g(x))$.  Because SO$(2)$ is a one-dimensional group,  we single out one angular coordinate, denoted by $\textrm{h}_{\varphi}$ and demand that the new  representation $(\textrm{h}_{\varphi},\mathbf{h})\in[-\pi,\pi]\times\mathbb{R}^M$ transforms equivariantly under  the SO$(2)$ group action as
\begin{align}
S_{\Delta\varphi}(\textrm{h}_{\varphi},\mathbf{h})=(\textrm{h}_{\varphi}+\Delta\varphi,\mathbf{h}).\label{Eq:group-action}
\end{align}
\begin{figure}
\includegraphics{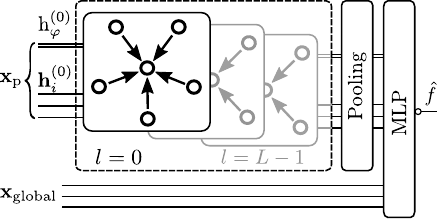}
\caption{\label{fig:network}Sketch of the network configuration as explained in the text. The equivariant feature is represented by a double line and the algorithms output by $\hat f$.}
\end{figure}
This group action applies a common shift $\Delta\varphi$ to the azimuthal coordinates $h_{\varphi}$ and to the azimuthal input features $\varphi_i$. The $M$ remaining features $\mathbf{h}$ of this data representation transform invariantly.  $\mathbf{h}$ will encode information related to the substructure of the jet, that are also invariant with respect to these rotations, while $h_{\varphi}$ can only contain information associated to quantities that transform equivariantly with the angle, effectively decoupling the two aspects.

Because the group manifold of SO$(2)$ is a circle,  we represent this coordinate in the complex plane as $e^{i h_{\varphi,i}}$ such that 
the message-passing relations
\begin{eqnarray}
\mathbf{h}^{(l+1)}_i&=&\sum_{j\in N(i)} \omega_j^{(N(i))}\boldsymbol{f}^{(l)}_{\textrm{h}}\left(\mathbf{h}^{(l)}_i, \mathbf{h}^{(l)}_j,h_{\varphi,i}^{(l)}-h_{\varphi,j}^{(l)}\right)\nonumber\\
e^{ih_{\varphi,i}^{(l+1)}}&=&e^{ih_{\varphi,i}^{(l)}+i\sum_{j\in N(i)}\omega_j^{(N(i))}f^{(l)}_\upphi\left(\mathbf{h}^{(l)}_i, \mathbf{h}^{(l)}_j,h_{\varphi,i}^{(l)}-h_{\varphi,j}^{(l)}\right)}\nonumber\\\label{Eq:equivariant-phi}
\end{eqnarray}
are manifestly equivariant under the transformation~(\ref{Eq:group-action}) if we restrict the inputs to the message-passing neural networks, $f^{(l)}_{\upphi}$ and $\boldsymbol{f}^{(l)}_{\textrm{h}}$, to group invariants. The distance function for $N(i)$ for $l>1$ can be generalized to $\mathbf{h}_i^2+\mathbf{h}_j^2-2\mathbf{h}_i\cdot\mathbf{h}_j\cos(h_{\varphi,i}-h_{\varphi,j})$.
Besides the features $\mathbf{h}$, invariant under $S$ by construction, the only other invariants are differences of $\varphi$ coordinates, such that Eq.~\ref{Eq:equivariant-phi} is general. Other dependencies are not allowed when IRC safety should be ensured, such that $\boldsymbol{x}_{\textrm{p,i}}$ reduces to $p_{i}^\mu$, and concretely for the inputs to the first iteration,
\begin{equation}
\begin{aligned}
\textrm{h}_{\varphi,i}^{(0)}=\varphi_i,\quad\mathbf{h}_i^{(0)}&=\Delta R_i,
\end{aligned}
\end{equation}
where $\Delta R_i$ is measured with respect to the jet axis.
Because the inputs to the neural networks $f_{\upphi}$ and $\boldsymbol{f}_{\textrm{h}}$ are the same, weights can be shared. In practice, a single neural network with $M^{(l)}+1$ outputs, simultaneously providing $f_{\upphi}$ and $\boldsymbol{f}_{\textrm{h}}$, has proven efficient.
After $L$ iterations, the global pooling is carried out with the energy-weighting in Eq.~\ref{Eq:energy-weighted-message}, but summing over all the constituents, i.e., $\mathcal{N}=\{\mathbf{x}_{\textrm{p}}\}_{i=1}^{N_{\textrm{p}}}$~\cite{Konar:2021zdg}. Together with the global features $\mathbf{x}_{\textrm{global}}$, the pooled IRC-safe and SO$(2)$ equivariant outputs are fed into a final MLP. A sketch of the whole construction is provided in Fig.~\ref{fig:network}.

Technically, the algorithm is implemented in \texttt{PyTorch}~\cite{pytorch} using the \texttt{PyTorch-Geometric}~(PyG) package~\cite{torch-geometric}.
Best performances were obtained with leaky ReLU activation functions with a slope parameter of 0.3.
For training, we used the \texttt{Adam} \texttt{PyTorch} optimizer with a learning rate of $10^{-3}$.
Other choices for the activation functions (e.g., {ReLU or sigmoid) and for the optimizer led to comparable performances, albeit with a slightly longer training time. This suggests that the algorithm performance is relatively robust. For the studies in the remainder of the paper, we found that a single iteration ($L=1$) with a message-passing neural network with two hidden layers of node size 20 and a read-out MLP with the same configuration performs sufficiently well. This is, therefore, our baseline configuration. Higher numbers of iterations train slower, with similar performance for the cases discussed below.

Finally, we note the possibility of breaking IRC safety by including extra information via the input features and still keeping the energy-weighting in Eq.~\ref{Eq:energy-weighted-message}. For example, the constituents charge could help resolve the parton-level ambiguity of the light quarks from the $\PW$~decay. While jet charge measurements were done by the ATLAS and CMS collaborations~\cite{ATLAS:2015rlw,CMS:2016lmd}, the benefit of non-IRC-safe information, including vertex-association quality or the probability of originating from pileup vertices, cannot be confidently assessed without access to a comprehensive model of systematic uncertainties. 

\section{Two toy scenarios}\label{sec:toy}
\begin{figure*}
\includegraphics[width=0.32\textwidth]{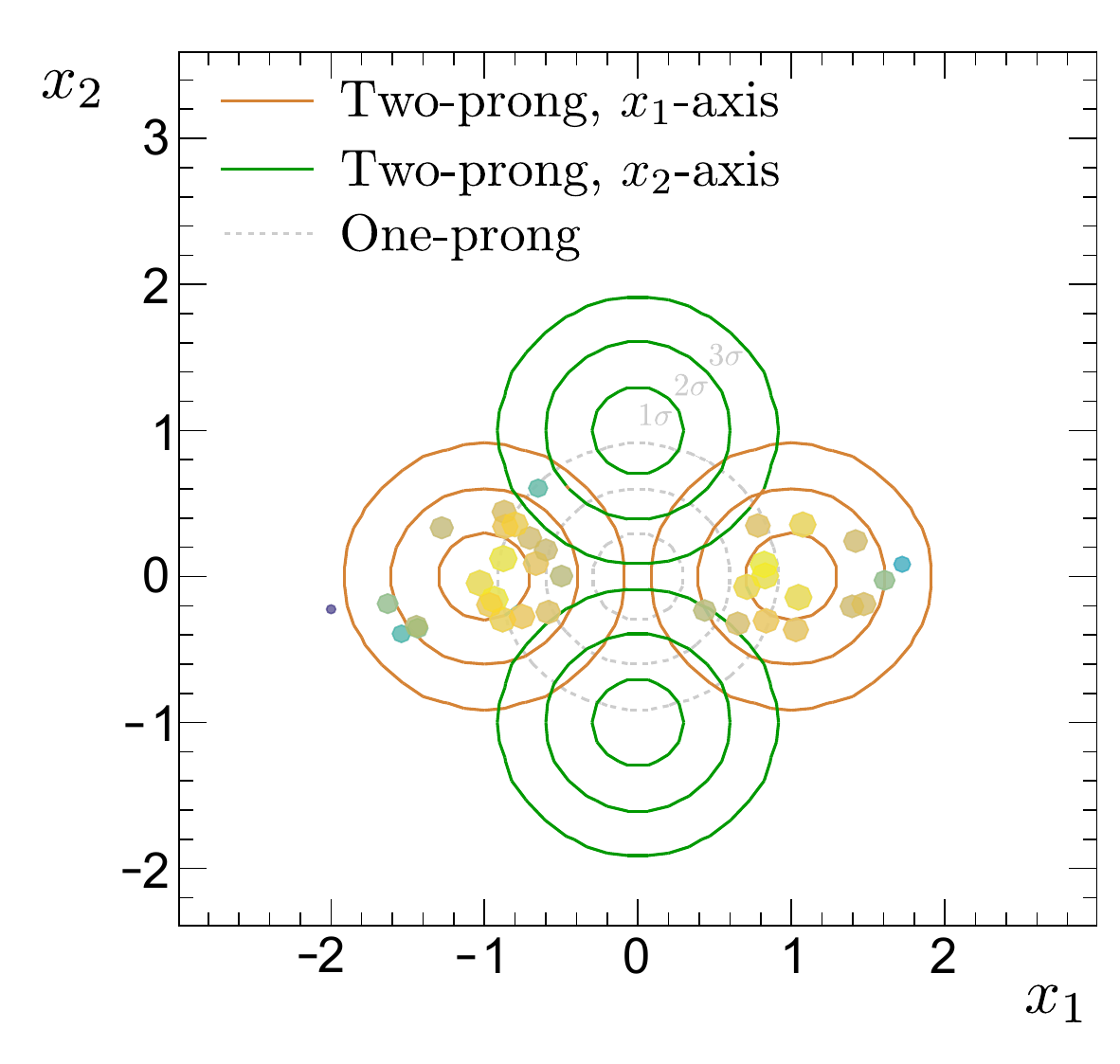}
\includegraphics[width=0.32\textwidth]{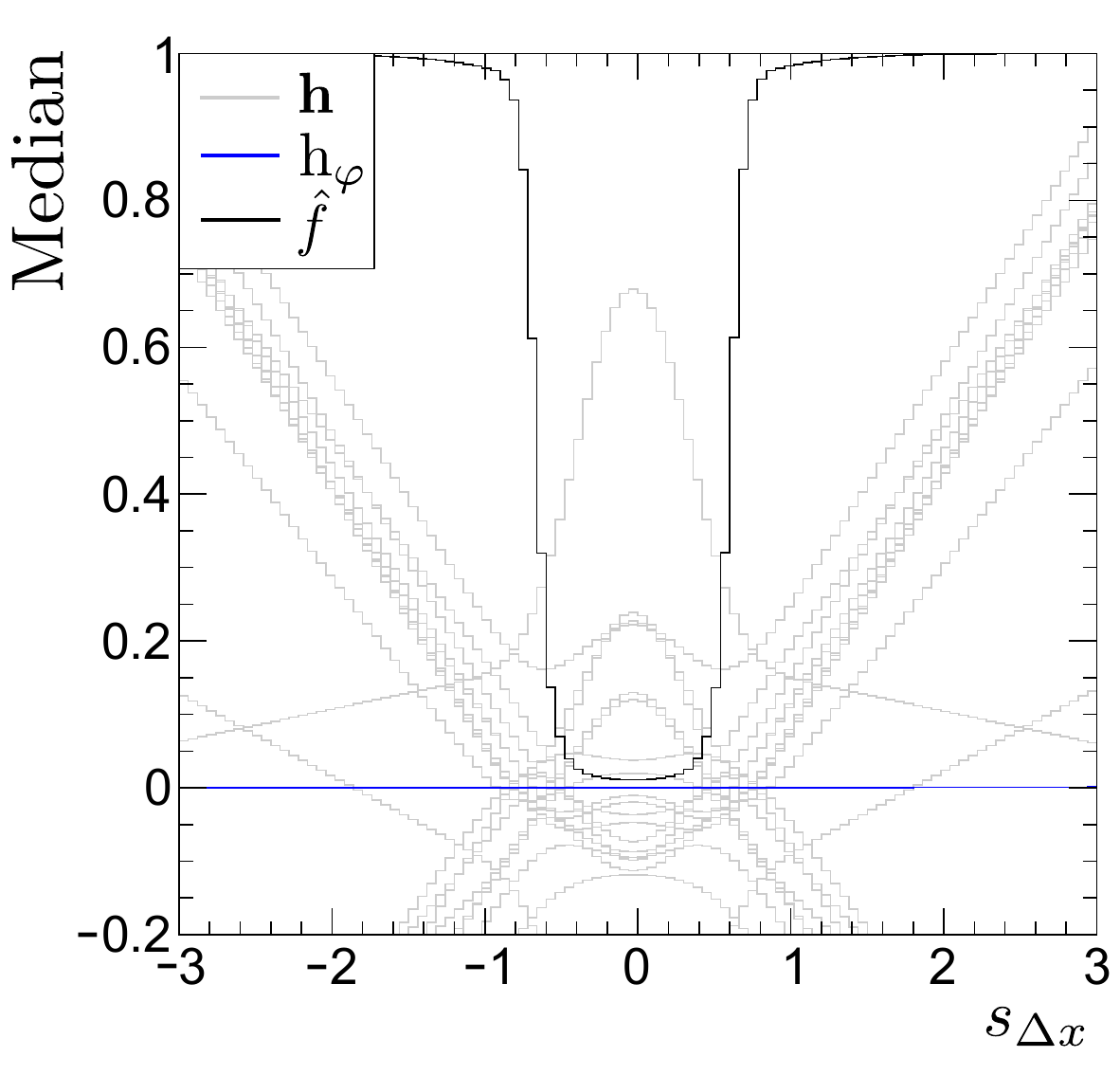}
\includegraphics[width=0.32\textwidth]{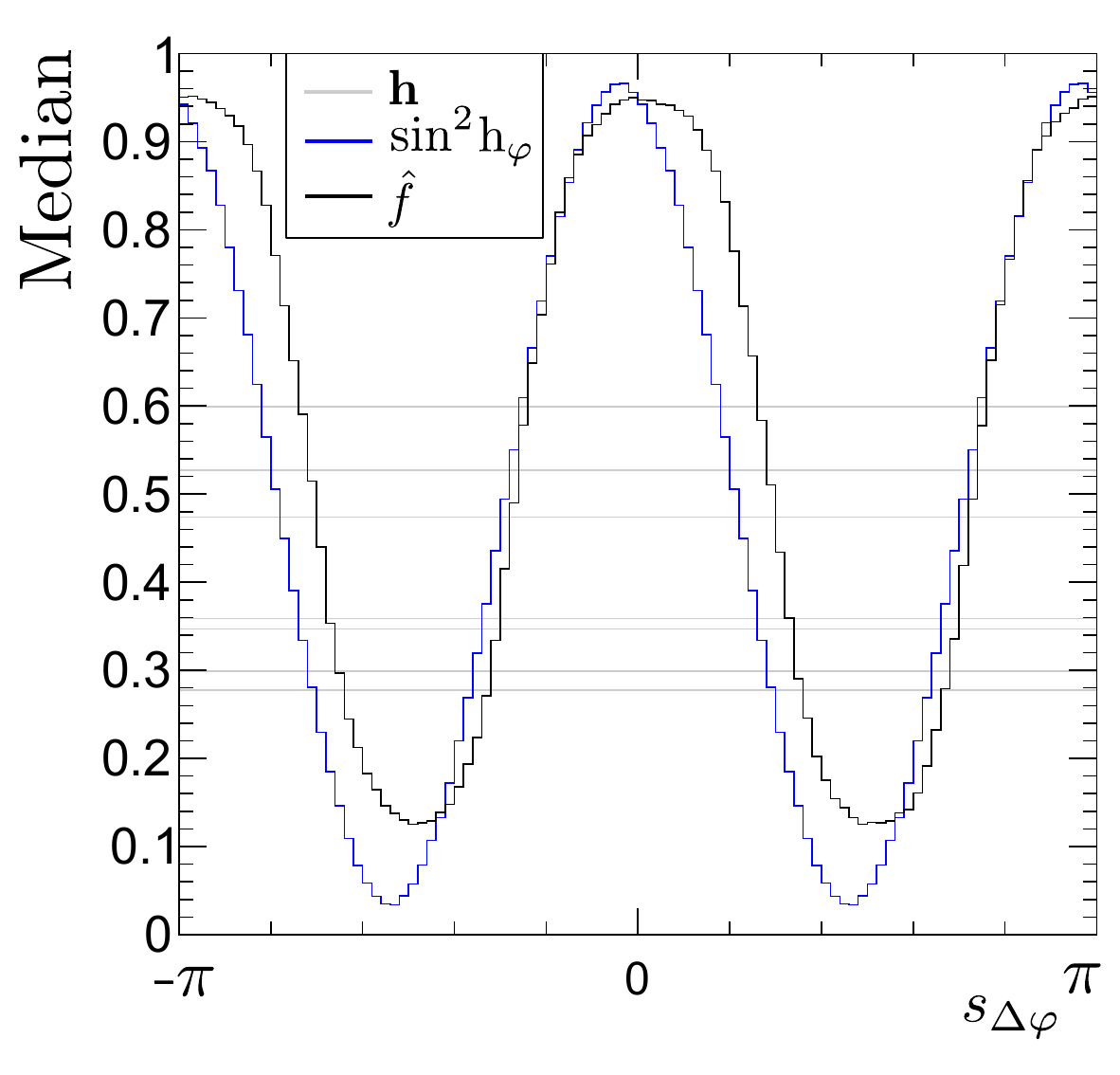}
\caption{\label{fig:toy-pdfs} Left: Contours of the probability density functions used in the toy example, along with an illustrative single event. Middle: Median of the MLP output score~(black) in scenario (A) as well as the medians of the 20 gNN outputs that serve as inputs to the redout MLP~(gray) in arbitrary units. The equivariant feature $h_{\gamma}$ is shown in blue. Right: Median of the MLP output score~(black) in scenario (B) as well as the medians of the first 7 pooled gNN outputs that serve as inputs to the redout MLP~(gray) in arbitrary units. The median of $\sin^2 \textrm{h}_\varphi$ is shown in blue. }
\end{figure*}
We illustrate the algorithms' basic properties using a  toy data set that aims to reflect a simple two-prong structure in a variable-length feature vector. We first let $\bar N_{\textrm{part}}=50$ be the mean number of constituents in a mock-up jet. For each training instance, we sample two Poisson random numbers $N_{1,2}$ with means $\bar N_{1,2}=\bar N_{\textrm{part}}/2$ such that each prong has, on average, the same number of constituents.
Subsequently, we draw $N_{1}$ and $N_{2}$ two-dimensional coordinates $\boldsymbol{x}_i=(x_{1},x_{2})_i$ from Normal distributions $\textrm{N}(\boldsymbol{\mu_1}, \boldsymbol{\Sigma})$ and $\textrm{N}(\boldsymbol{\mu_2}, \boldsymbol{\Sigma})$, respectively. We choose $\boldsymbol{\Sigma}=\textrm{Diag}(0.3,0.3)$. For the constituents' momenta $p_{\textrm{T},i}$, we draw random numbers from a log-normal distribution $\textrm{lnN}(1,0.2)$. We scale the $p_{\textrm{T},i}$ by the likelihood of the constituent's 2D-coordinate in the Normal distribution it was obtained from, that is $N(\boldsymbol{x}_i|\boldsymbol{\mu}_{1,2},\boldsymbol{\Sigma})$. Because we sampled the $\mathbf{x}$ from Normal distributions, each prong will likely have high-momentum particles close to the location parameters $\boldsymbol{\mu_{1,2}}$. The sum of all constituent momenta, taken over both prongs, is finally scaled to 100 such that the mockup jet's total momentum is not random. During training, there is no information on the constituents' origin; each instance in the training set $\mathcal{D}$ is given solely by a total of $N_j=N_{j,1}+N_{j,2}$ constituents $\{p_{\textrm{T},i},\boldsymbol{x}_i\}_{i=1}^{N_j}$ and a binary truth label $y_j$.
There are no global event features in this study. The loss function, for simplicity, is the mean-squared-error
\begin{align}
L=\sum_{\{p,\boldsymbol{x}\}_j, y_j\in\mathcal{D}}\Bigg(\hat f(\{\boldsymbol{x}_i,p_i\}_{i=1}^{N_j})-y_j\Bigg)^2.
\end{align}
We have also tested the binary cross-entropy loss function and found no change in the performance.

In a scenario (A) we attempt a simple classification task between a two-prong signal category $\boldsymbol{\mu}^{\textrm{sig}}_{1,2}=(\pm1,0)$ and a background category $\boldsymbol{\mu}^{\textrm{bkg}}_{1,2}=(0,0)$, the latter equivalent to a single prong with a mean of $\bar N_{\textrm{part}}$ constituents. Contour lines of the probability density function~(pdf) at $1\sigma$, $2\sigma$, and $3\sigma$ levels alongside the constituents of a  single illustrative signal event are shown in Fig.~\ref{fig:toy-pdfs}~(left).
Training this classifier on $10^5$ pseudo-events for both categories leads to a background efficiency of approx. $1\%$ for a signal efficiency of $99\%$. It is more interesting to study the behavior of the gNN output features before they enter the final MLP for different types of input. For this purpose, we exploit the simplicity of our toy data set: a sample of signal events is, by construction, indistinguishable from a background sample when the constituents are shifted by an amount $\mathbf{x}_i\rightarrow\mathbf{x}_i+\Delta\mathbf{x}_i$ with $\Delta\mathbf{x}_i=-\boldsymbol{\mu}_{1,2}$, depending on whether constituent $i$ came from prong 1 or prong 2. For illustration, we generate $10^4$ independent test events and shift the constituents by $\Delta \mathbf{x}_i=-\boldsymbol{\mu}_{1}+ (s_{\Delta x},0 )$ for prong 1 and $\Delta \mathbf{x}_i=-\boldsymbol{\mu}_{2}- (s_{\Delta x},0 )$. In this construction, the test sample is background-like for $s_{\Delta x}=0$ and signal-like for $s_{\Delta x}=1$. In Fig.~\ref{fig:toy-pdfs}~(middle) we show the median of the output score and the medians of the pooled gNN outputs before they enter the readout MLP.  It is evident that the readout MLP learns to distinguish signal events from background events using the 20-dimensional internal representation $\mathbf{h}$. The insignificance of the equivariant feature is expected, as there is no relevant azimuthal dependence of the prong structure in this classification task.

The situation is very different in scenario (B), where we train a classifier to separate between a signal with $\boldsymbol{\mu}^{\textrm{sig}}_{1,2}=(\pm 1,0)$ from a background category with $\boldsymbol{\mu}^{\textrm{bkg}}_{1,2}=(0, \pm 1)$. This time, the signal has a prong structure along the $x_1$-axis while the background has a prong structure along the $x_2$ axis. 
In Fig.~\ref{fig:toy-pdfs}~(right) we show similar results as before, but this time for a signal sample that is rotated around the origin by an angle $s_{\Delta\varphi}$. Therefore, $s_{\Delta\varphi}=0$ or $s_{\Delta\varphi}=\pm\pi$ correspond to a signal sample, while $s_{\Delta\varphi}=\pm \frac{\pi}{2}$ corresponds to a background sample. We first note from Fig.~\ref{fig:toy-pdfs}~(right) that the internal features do not vary at all with the angle of rotation $s_{\Delta\varphi}$. This manifests the invariance of $\mathbf{h}$ under the operation $S_{\Delta\varphi}$ in Eq.~\ref{Eq:equivariant-phi}. Secondly, we note that there is a reflection ambiguity in the toy setup that mimics a similar reflection ambiguity of the light quarks in the $\PW$~boson decay. The probability density function is symmetric under the exchange of $x_1$ and $x_2$; however, the value of $h_{\varphi}$ changes by $\pi$ under this operation. Therefore, we investigate a function of the equivariant feature which is invariant under this reflection. The simplest such function is $\sin^2 h_{\phi}$. The median of this value in the test data set is shown in blue color in Fig.~\ref{fig:toy-pdfs}~(right) as a function of $s_{\Delta \varphi}$. It transforms with the rotation of inputs such that the events in the test sample are identified as signal events for rotations that are close to even multiples of $\pi/2$ and as background when the rotation is close to odd multiples of $\pi/2$. The modulations of $\sin^2h_{\varphi}$ and the classifier output are not in phase, but that only reflects the networks' freedom in the internal representation. 

In summary, the trained algorithm correctly reflects the basic properties of the classification task in both scenarios. 

\section{Learning from the WZ process}\label{sec:WZEFT-results}
We now showcase the performance of the algorithm in a more realistic setup using a sample of simulated $\PW\PZ$ events in a semileptonic decay chain. The $\Pp\Pp \to \PW(\to\Pq\bar\Pq)\PZ(\to\ell\bar\ell)$ events are generated with \texttt{MADGRAPH5\_aMC@NLO}~v2.65 event generator~\cite{Alwall:2014hca} in the leading-order configuration with one extra parton in the hard-scatter process. 
The simulation sample uses the NNPDF3.1 PDF set~\cite{Ball_2017}.
The renormalization and factorization scales are not kept fixed, and their values are the default in \texttt{MADGRAPH5},
namely the transverse mass of the $2 \rightarrow 2$ system resulting from $k_{\textrm{T}}$ clustering.
A requirement of \mbox{$H_\textrm{T}>300\,$GeV} selects events with a sufficiently boosted $\PW$ boson candidate, where $H_{\textrm{T}}$ is the scalar sum of transverse final-state parton momenta.  
 We use the \texttt{SMEFTSim}~v3.0~\cite{Brivio:2020onw} model with single SMEFT operator insertions to simulate an event sample at $C_{\PW} = C_{\widetilde\PW} = 1$ in order to ensure the kinematic phase space is well populated under the SM and SMEFT scenarios. 
The \texttt{PYTHIA}~v8.24~\cite{Sjostrand:2014zea} package is used to simulate the parton shower and hadronization.  
The matching between matrix element calculation in \texttt{MADGRAPH5} and \texttt{PYTHIA} parton shower model is performed following the MLM~\cite{Alwall:2014hca} prescription.
The detector response is emulated using \texttt{DELPHES}~v3.5.0~\cite{deFavereau:2013fsa} with the ATLAS card, such that the setup is equivalent to the top quark reference data set described in Ref.~\cite{kasieczka_gregor_2019_2603256}.

Since the SMEFT effects induce a modulation in the distribution of $\varphi_{\textrm{decay}}$, we first use the algorithm to regress in this variable from the jet's constituents. Analogously to Sec.\ref{sec:toy}, the  observed lab-frame features of a training event $j$ are given by $\mathbf{x}_j=\{p_{\textrm{T},i},\varphi_i,\Delta R_i\}_{i=1}^{N_j}$ where the $\varphi_i$ and $\Delta R_i$   are measured with respect the jet's axis and $N_j$ denotes the number of jet constituents. The parton-level $\varphi_{j,\textrm{decay}}$ is the regression target. 
We consider only events with a reconstructed AK8 jet with \mbox{$p_{\textrm{T}} > 500\,$GeV} and we do not use any global event information. We train the algorithm on 80\% of the \PW\PZ~ data set by minimizing the loss function
\begin{align}
L=\sum_{\mathbf{x}_j\in\mathcal{D}_{\textrm{sim}}}\sin^2 \Big(\hat f(\mathbf{x}_j)-\varphi_{j,\textrm{decay}}\Big).
\end{align}
This choice, in particular the sine function, implements invariance to the reflection ambiguity, described in Sec.~\ref{sec:toy}, at the level of the loss function: the data does not provide information to distinguish constituents originating from up- from down-type quarks and therefore will be invariant under transformations $\varphi_{j,\textrm{decay}} \to \pi-\varphi_{j,\textrm{decay}}$. The sine removes the ambiguity and this inductive bias improves the speed of convergence. It is, however, not needed in principle. We have tested that the network also converges with a standard mean-squared error loss or, e.g., with a piecewise linear function with the same symmetry properties as the sine function. Figure~\ref{fig:decay_plane_regression_2d} shows the two-dimensional distribution of the true and the regressed $\varphi_{\textrm{decay}}$ in the remaining 20\% \PW\PZ~data set, indicating a sensible behavior of the algorithm.

\begin{figure}
\includegraphics[width=\linewidth]{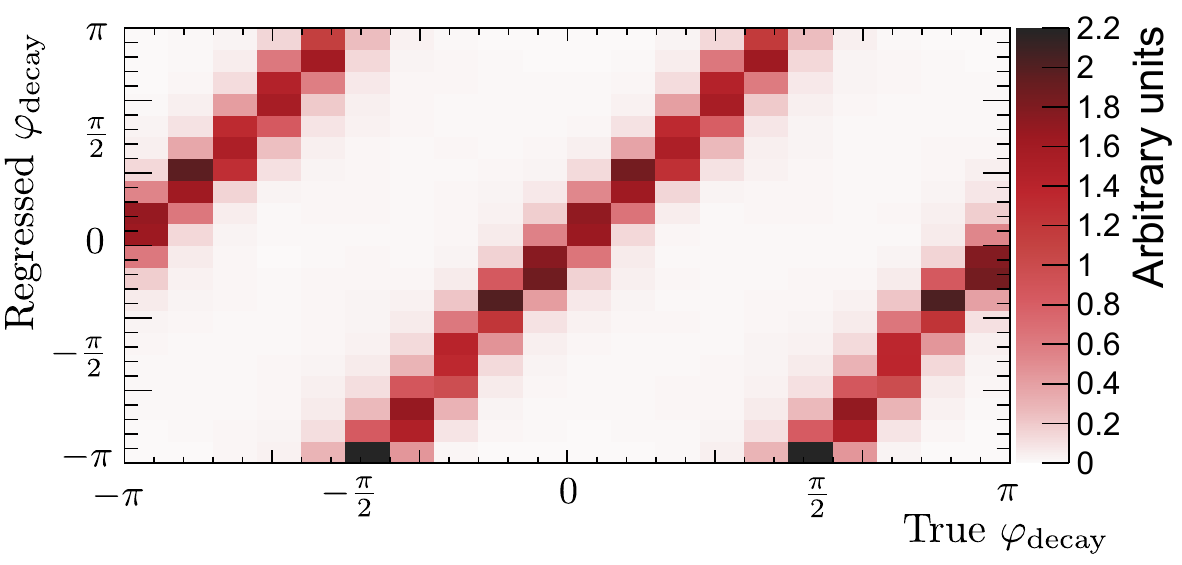}
\caption{Two-dimensional distribution of the true and regressed  angle $\varphi_{\textrm{decay}}$ in the test dataset.}
\label{fig:decay_plane_regression_2d}
\end{figure}

We now turn to extracting SMEFT sensitivity, considering events with a Z boson candidate constructed from its decay products, and a reconstructed jet, both with \mbox{$p_{\textrm{T}} > 300\,$GeV}. For predicting SMEFT effects, we use a variant of the morphing technique~\cite{Brehmer:2019xox} to obtain weighted predictions in the parameter space spanned by the Wilson coefficients. To this end, \texttt{MADGRAPH5} with the \texttt{MADWEIGHT}~\cite{Artoisenet:2010cn} module is used to compute a per-event weight for a sufficiently large number of SMEFT parameter points $\mathbf{C}=\Big(C_\PW,C_{\widetilde\PW}\Big)$. Because second-order polynomials accurately describe the SMEFT matrix elements with a single operator insertion, a small number of such evaluations is sufficient to determine the coefficients of this polynomial, denoted by $\omega_j(\mathbf{C})$ for an event with a label $j$. We choose an overall normalization of the event weights such that at the SM parameter point $\mathbf{C}=0$ we have $\sum_j\omega_j(0)=\mathcal{L}\,\sigma^{\PW\PZ}(\textrm{SM})$.
It follows that the SMEFT prediction for any Poisson yield $\lambda(\mathbf{C})$ in a phase space volume $\Delta\mathbf{x}$, defined in terms of the observed features $\mathbf{x}$, is given by the sum of the polynomial per-event weights
\begin{align}
\lambda(\mathbf{C})&=\mathcal{L}\int_{\Delta\mathbf{x}}\textrm{d}\mathbf{x}\frac{\textrm{d}\sigma(\mathbf{C})}{\textrm{d}\mathbf{x}}=\mathcal{L}\sigma(\mathbf{C})\int_{\Delta\mathbf{x}}
\textrm{d}\mathbf{x}\,p(\mathbf{x}|\mathbf{C})\nonumber\\
&=\mathcal{L}\sigma(\mathbf{C})\int_{\Delta\mathbf{x}}
\textrm{d}\mathbf{x}\int\textrm{d}\mathbf{z}\,p(\mathbf{x},\mathbf{z}|\mathbf{C})\approx\sum_{\mathbf{x}_j\in\Delta\mathbf{x}}\omega_j(\mathbf{C}).
\end{align}
The second expression gives the yield in terms of the differential cross-section, the third relates the cross-section to the detector-level pdf $p(\mathbf{x}|C)$, the fourth introduces the joint pdf of the observation $\mathbf{x}$ and the latent features $\mathbf{z}$, and the last term, finally, is the Monte-Carlo approximation from a simulated sample $\mathcal{D}_{\textrm{sim}}=\{\omega_j(\mathbf{C}),\mathbf{x}_j,\mathbf{z}_j\}_{j=1}^{N_{\textrm{sim}}}$. 
Because Monte-Carlo simulation is a sampling of the joint $(\mathbf{x},\mathbf{z})$-space space~\cite{Brehmer:2018eca}, the ratio of two polynomial event weights is expressed in terms of the ratio of the joint likelihood,
\begin{align}
\frac{\omega_j(\mathbf{C})}{\omega_j(0)}=\frac{\sigma(\mathbf{C})}{\sigma(\textrm{SM})}\frac{p(\mathbf{x}_j,\mathbf{z}_j|\mathbf{C})\phantom{C}}{p(\mathbf{x}_j,\mathbf{z}_j|\textrm{SM})}
.\label{eq:weight-ratio}
\end{align}
Here, the first factor on the r.h.s. accounts for the dependence of the total cross-section on the Wilson coefficients. 
The simulation-based inference technique we use for learning the SM-SMEFT interference is based on Refs.~\cite{Cranmer:2015bka,Brehmer:2018kdj,Brehmer:2018eca,Brehmer:2018hga,Brehmer:2019xox,Brehmer:2019gmn} and similar to the SALLY method~\cite{Brehmer:2018kdj}. It capitalizes on the fact that the joint likelihood ratio is tractable, i.e., simulation allows evaluating Eq.~\ref{eq:weight-ratio} as a function of $\mathbf{C}$. 

\begin{figure*}
\includegraphics[width=0.33\textwidth]{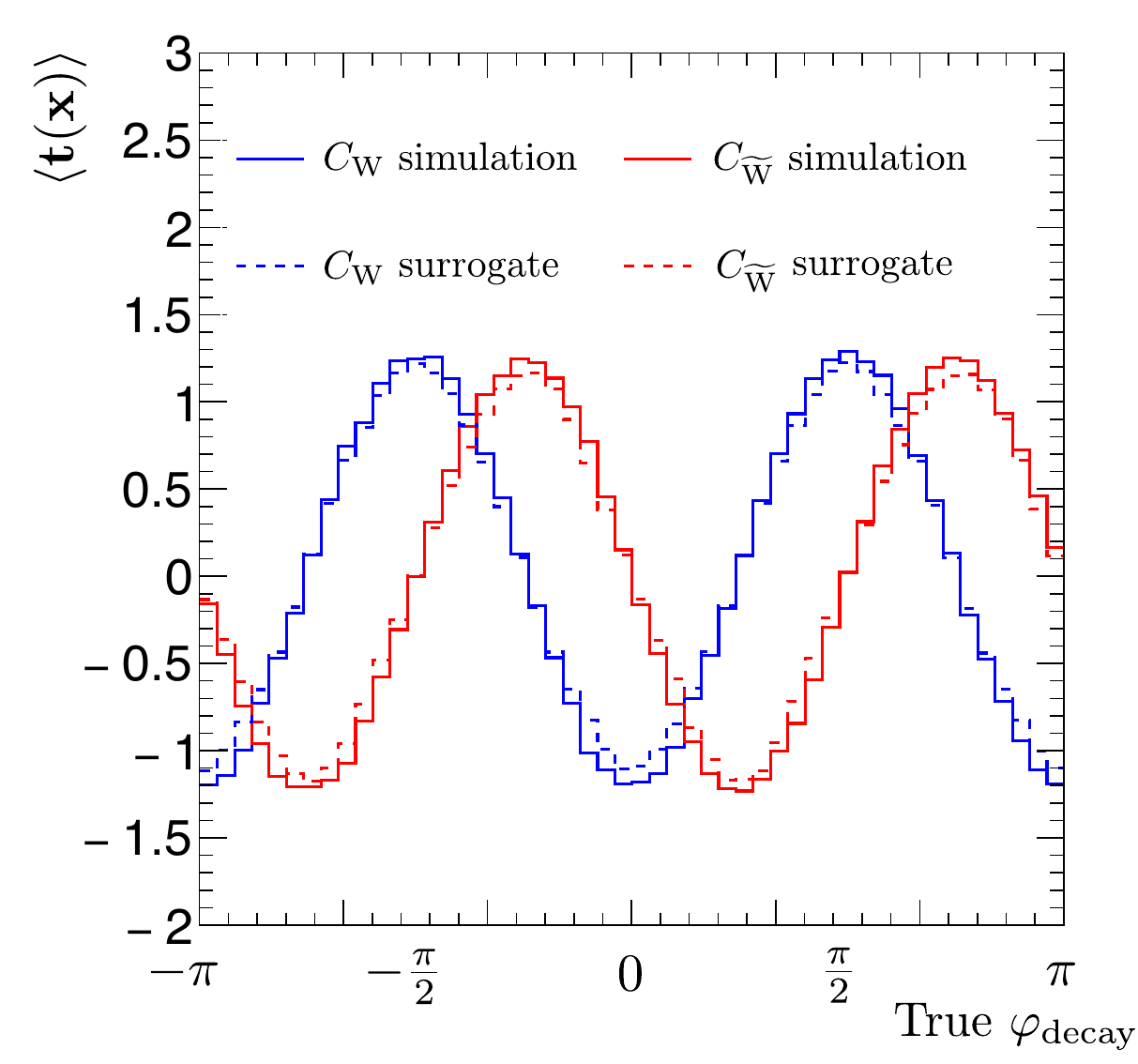}
\includegraphics[width=0.32\textwidth]{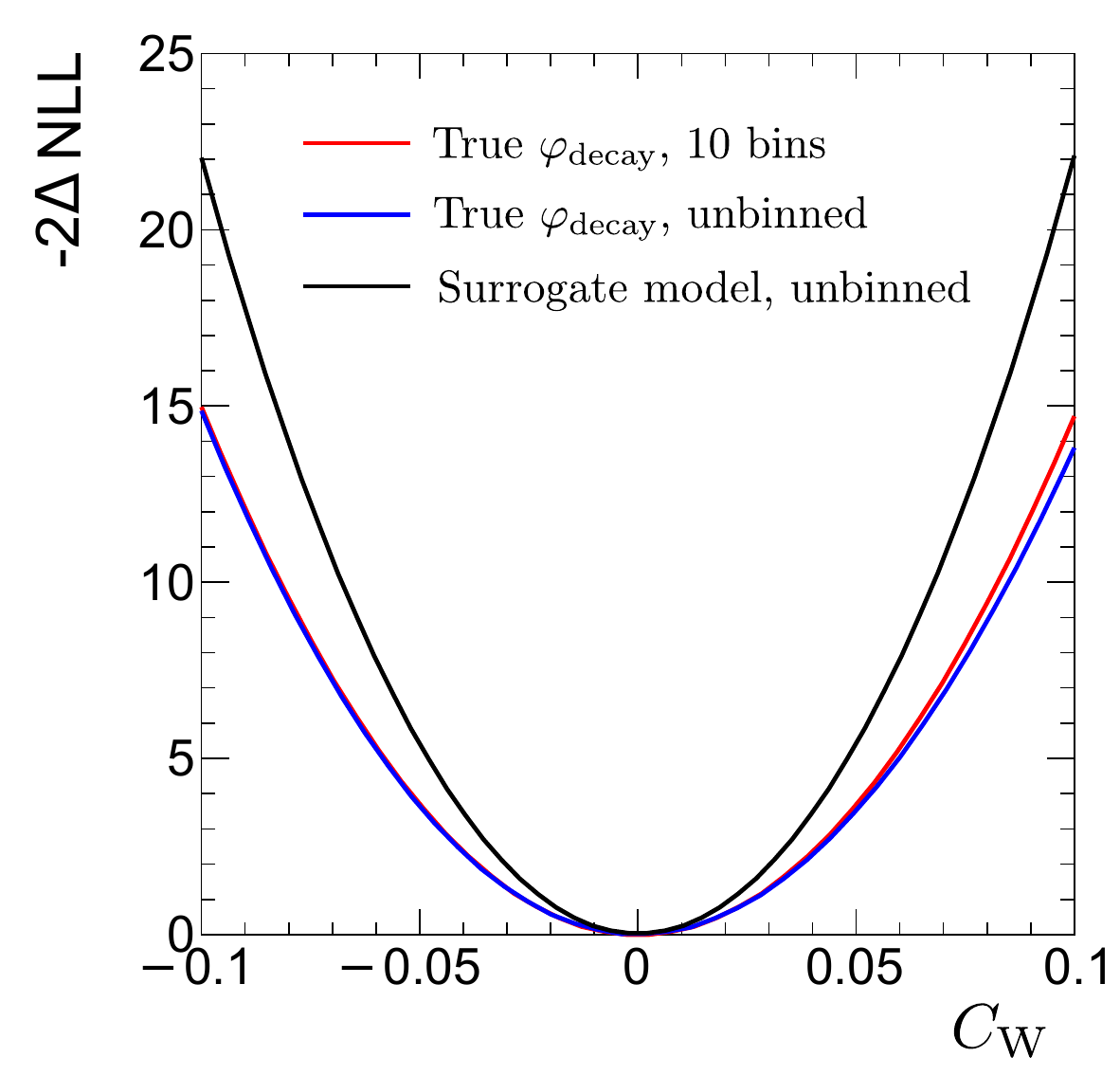}
\includegraphics[width=0.32\textwidth]{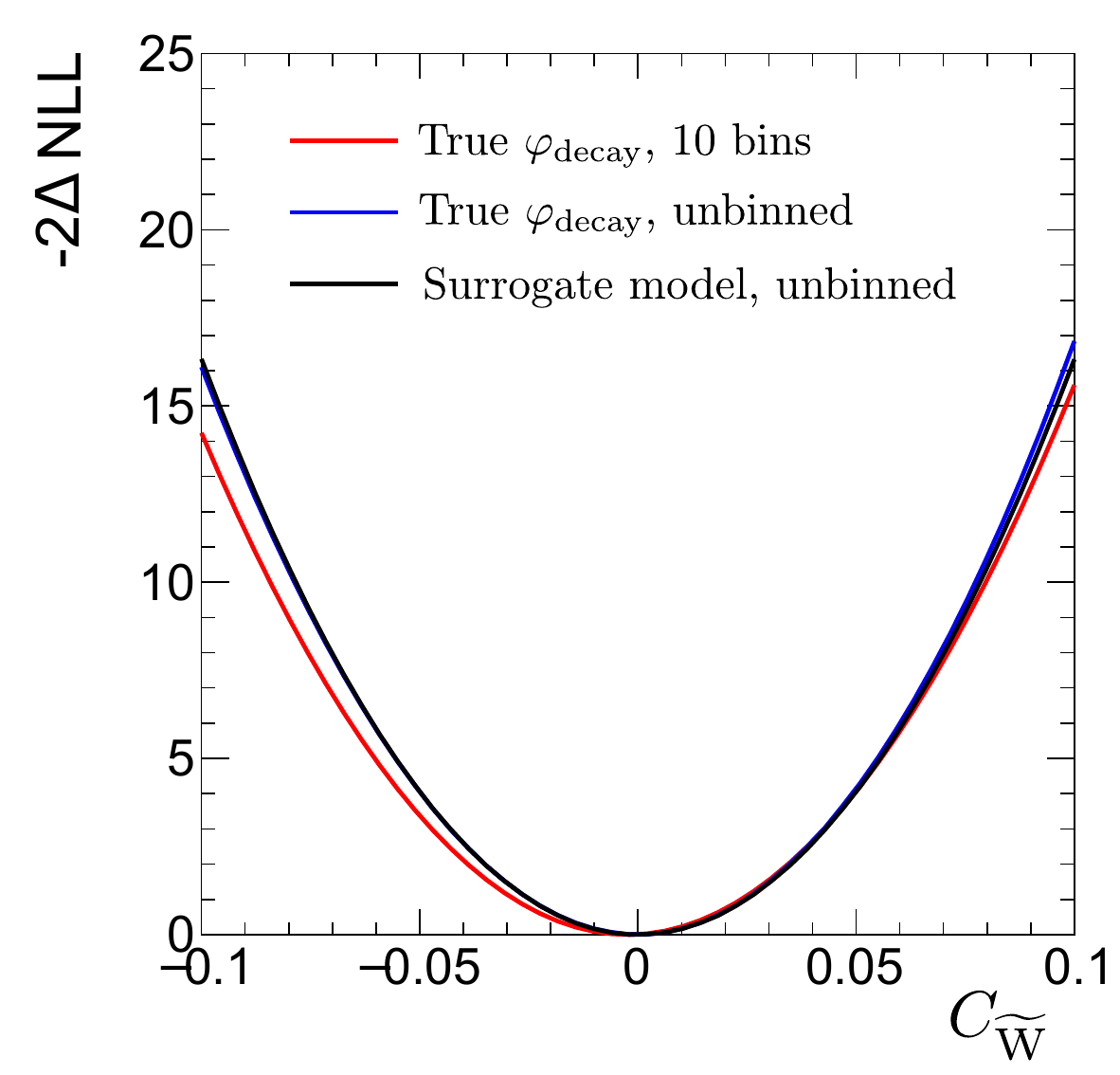}
\caption{\label{fig:closure} Left: Mean score as a function of $\varphi_{\textrm{decay}}$, obtained with the true model and the regressed surrogate model. Center and right: likelihood scan as a function of $C_{\PW}$ (center) and $C_{\widetilde\PW}$ (right) using different likelihood models, assuming 2000 observed events. The surrogate model described in this paper is compared with a MLP that takes as a only input the $\varphi_{\textrm{decay}}$ and the likelihood function of a counting experiment considering 10 bins in $\varphi_{\textrm{decay}}$. }
\end{figure*}
Because our targeted effect is linear in $\mathbf{C}$,
we are interested in the model's behavior near the SM where the score vector 
\begin{align}
\mathbf{t}(\mathbf{x})=\nabla_{\mathbf{C}}\log p(\mathbf{x}|\mathbf{C})|_{\mathbf{C}=\textrm{SM}}=\frac{\nabla_{\mathbf{C}}\;p(\mathbf{x}|\mathbf{C})}{p(\mathbf{x}|\textrm{SM})}\Bigg|_{\mathbf{C}=\textrm{SM}}\nonumber\\\label{eq:score}
\end{align}is a sufficient statistic~\cite{Brehmer:2018eca}. It provides a locally optimal observable, i.e., extracts the maximum amount of information from the training data~\cite{Brehmer:2018eca}. At first sight, it appears problematic that our quantity of interest in Eq.~\ref{eq:score} is a ratio of detector-level likelihoods, with separate implicit latent-space integrations in the numerator and denominator, while in Eq.~\ref{eq:weight-ratio}, the quantity available in simulation, there appears a ratio of latent-space pdfs without that integration. Following  Refs.~\cite{Cranmer:2015bka,Brehmer:2018kdj}, we can nevertheless learn the detector-level score from expanding the log-derivative of Eq.~\ref{eq:weight-ratio} to linear order, 
\begin{align}
    \mathbf{t}(\mathbf{x}_j,\mathbf{z}_j)=\nabla_{\mathbf{C}}\log\frac{\omega_j(\mathbf{C})}{\sigma(\mathbf{C})}\Bigg|_{\mathbf{C}=\textrm{SM}}.\label{eq:empirical-joint-score}
\end{align}
This quantity is tractable for simulated events. 
It is then straightforward to show~\cite{Cranmer:2015bka} that the analytic loss function
\begin{align}
    L = \int\textrm{d}\mathbf{x}\,\textrm{d}\mathbf{z} \, p(\mathbf{x},\mathbf{z}|\textrm{SM})\left(\mathbf{t}(\mathbf{x},\mathbf{z})-\hat {\boldsymbol{f}}(\mathbf{x})\right)^2\label{eq:loss-SBI-analytic}
\end{align}
for an infinitely expressive $\hat{\boldsymbol{f}}$,  depending on the observable $\mathbf{x}$ but not on the latent $\mathbf{z}$, provides a minimum $\boldsymbol{f}^{\ast}(\mathbf{x})=\mathbf{t}(\mathbf{x})$ as defined in Eq.~\ref{eq:score}. Our empirical loss function is, therefore, the empirical version of Eq.~\ref{eq:loss-SBI-analytic}, given by
\begin{align}
    L = \sum_{\{\omega_j(\mathbf{C}),\mathbf{x}_j,\mathbf{z}_j\}\in\mathcal{D}_{\textrm{sim}}}\omega_j(\textrm{SM})\left(\mathbf{t}(\mathbf{x}_j,\mathbf{z}_j)-\hat{\boldsymbol{f}}(\mathbf{x}_j)\right)^2\label{eq:loss-SBI-empirical}
\end{align}
where we make use of Eq.~\ref{eq:empirical-joint-score}. All its ingredients are available in the simulation. Finally, we can remove $\sigma(\mathbf{C})$ in Eq.~\ref{eq:empirical-joint-score} if we use $\hat{\boldsymbol{f}}$ only as a discriminator in a hypothesis test. The total cross-section has no impact on its discriminative power as this term only provides a constant shift to the predicted value, common to all events. 

Minimizing this loss function over $3 \cdot10^5$ training events provides a surrogate of the detector-level two-component score vector of $C_{\PW}$ and $C_{\widetilde\PW}$. The distance parameter $\Delta R$ in the gNN is set to the value of 0.4.
The same network configuration as in Sec.~\ref{sec:toy} is used, with little observed dependence on the MLP configuration. The detector-level transverse momenta of the leptonic and hadronic bosons show mild dependence on $C_{\PW}$  at the linear level and are, therefore, used as global features for the $C_{\PW}$ component, while we do not use global features for the $C_{\widetilde\PW}$ component.

We show the result of the training in Fig.~\ref{fig:closure}~(left) on a statistically independent test data set. The detector-level score exhibits the expected sinusoidal modulations as a function of the true decay plane angle. The gNN very accurately reproduces this dependence, showing that the gNN can indeed recover SMEFT effects in the angular radiation patterns. 

In Fig.~\ref{fig:closure}~(middle, right) we perform one-dimensional likelihood scans of our surrogate model normalized to 2000 expected events. Without incorporating systematic uncertainties, this procedure will lead to optimistic results. Nevertheless, comparing the surrogate model's performance to the truth level is enlightening. The negative log-likelihood (NLL) ratio is shown for three different test statistics. First, the true decay-plane angle $\varphi_{\textrm{decay}}$ shows similar performance in a binned Poisson and as an unbinned test statistic, suggesting that a binned analysis in the regressed $\varphi_{\textrm{decay}}$ may have good sensitivity. Second, the unbinned surrogate model performs well for both, the $C_{\PW}$ and $C_{\widetilde{\PW}}$ Wilson coefficients. 
Despite the complexity of the hadronic final state, the gNN thus manages to extract the leading linear SMEFT dependence from the boosted jet's constituents.
In the case of $C_{\PW}$, the surrogate model also profits from the SMEFT dependence of the global event features, increasing the sensitivity. This showcases the ability of the algorithm to combine information from the radiation patterns and observables sensitive to energy growth. For the $C_{\widetilde{\PW}}$ coefficient, the energy growth is much smaller and $\varphi_{\textrm{decay}}$ dominates the sensitivity. 

In combination, the results show that our gNN can extract the linear SMEFT dependence from the decay plane angle of a hadronically decaying \PW~boson, unlocking the large hadronic branching fractions for future analyses in search of new physics.

\section{Conclusions and Outlook}\label{sec:outlook}


In this paper, we have constructed an IRC-safe and rotation-equivariant gNN, whose main input is the variable-length list of the particle constituents of a jet produced in the hadronic decay of a boosted massive particle. The algorithm is able to extract information on the angular orientation of the radiation patterns present in the jet, decoupled from other aspects of the jet substructure thanks to the equivariance of the network.

We have investigated the main features in simple toy studies and applied the algorithm to linear SMEFT effects in semileptonic final states of the $\PW\PZ$ process at the LHC.
It learns a surrogate of the score vector, thus providing an optimal observable for small deviations from the SM. We have shown that the score is well reproduced as a function of the true decay plane angle, thus fully extracting its SMEFT sensitivity.
Moreover, we incorporate information from additional observables encoding effects from energy growth, boosting the sensitivity towards the theoretical optimum.

The algorithm allows the large hadronic branching fractions to be utilized in future SMEFT analyses.
It is also suitably general to be useful in a variety of applications, potentially including three-prong decays of boosted top quarks or semileptonic final states of vector boson scattering at the LHC.

\begin{acknowledgments}
S. S\'anchez Cruz's research work has been funded by the Postdoc Grant program of the University of Z\"urich (reference FK-23-132). D. Schwarz's research work has been funded by the Austrian Science Fund~(FWF, grant P33771). 

\end{acknowledgments}

\bibliographystyle{lucas_unsrt}
\bibliography{biblio}

\end{document}